\documentclass{article}

\font\sqi=cmssq8
\def\DR{\rm I\kern-1.45pt\rm R}
\def\DC{\kern2pt {\hbox{\sqi I}}\kern-4.2pt\rm C}
\newcommand{\ben}{\begin{enumerate}}
\newcommand{\een}{\end{enumerate}}
\newcommand{\beq}{\begin{equation}}
\newcommand{\eeq}{\end{equation}}
\newcommand{\bse}{\begin{subequation}}
\newcommand{\ese}{\end{subequation}}
\newcommand{\bea}{\begin{eqnarray}}
\newcommand{\eea}{\end{eqnarray}}
\newcommand{\bc}{\begin{center}}
\newcommand{\ec}{\end{center}}

\newcommand{\ch}{{\tt h}}

\def\DH{\rm I\kern-1.5pt\rm H\kern-1.5pt\rm I}
\begin{document}
\begin{center}
{\large\bf  Four-dimensional Hall mechanics as a particle on $\DC P^3$
} \\
\vspace{0.5 cm}
{\large Stefano Bellucci$^1$, Pierre-Yves Casteill$^1$
and   Armen Nersessian$^{2,3}$ }
\end{center}
{\it $^1$ INFN-Laboratori Nazionali di Frascati,
 P.O. Box 13, I-00044, Italy\\
$^2$ Yerevan State University, Alex  Manoogian St., 1, Yerevan,
375025, Armenia\\
$^3$ Yerevan Physics Institute, Alikhanian Brothers St., 2, Yerevan, 375036,
 Armenia}
\begin{abstract}
In order to establish an explicit connection between four-dimensional Hall
effect on $S^4$ and six-dimensional Hall effect on $\DC P^3$,
we perform the Hamiltonian reduction
of a particle moving on $\DC P^3$  in a constant magnetic field to
the four-dimensional Hall mechanics (i.e. a particle on
 $S^4$ in a $SU(2)$ instanton field).
This reduction corresponds to fixing the isospin of the latter
system.
\end{abstract}
\setcounter{equation}0
Recently, Zhang and Hu found an interesting
four-dimensional generalization of quantum Hall effect
based on the quantum mechanics of non-interacting  particles
moving on a four-dimensional
sphere in the $SU(2)$ instanton field,
 with effective three-dimensional edge dynamics \cite{4h}.
The four-dimensional Hall effect inherits
many properties of its conventional, two-dimensional counterpart \cite{2h}
The most interesting property of four-dimensional
Hall effect is that the ground state is separated from the
 excited states  by a finite energy gap,
 while  the density correlation functions decay gaussianly.
This theory was analysed from many viewpoints
and  extended in various directions \cite{Fabinger}-\cite{octonions}.
The close similarity of the four-dimensional Hall effect
by Zhang and Hu  with the conventional Hall effect is due to their connection
with the first Hopf map, $S^3/U(1)=\DC P^1\cong S^2$ and the second Hopf map,
 $S^7/SU(2)=\DH P^1\cong S^4$, respectively.
In fact, the four dimensional Hall
effect by Zhang and Hu could be viewed as a ``quaternionic''
 analog of conventional quantum Hall effect
(we wish to mention also the recent paper \cite{octonions}, where the
 ``octonionic"  Hall effect  on $S^8 $ has been suggested,
 based in the last Hopf map, $S^{15}/S^7=S^8$).
In order to obtain a reasonable thermodynamic
 limit with a finite spatial density of particles,
one has to consider very large  $SU(2)$ representations.
 In that case each particle has an infinite number
 of $SU(2)$ internal degrees of freedom. This yields the conclusion that the  spectrum of
 edge excitations contains massless particles of all spins.
 Karabali and Nair  avoided
 the introduction of an infinite number of internal  degrees of  freedom,
  suggesting a four-dimensional Hall
 effect based on the Landau problem on
the complex projective space $\DC P^2$, as well as its
$2N-$dimensional generalization
 on $\DC P^N$ \cite{kn}, with effective $2N-1$ edge dynamics.
Flat limits of the two pictures were considered in details
 by Elvang and Polchinski \cite{polchinski}.
In both variants a key role is played by the
presence of a instanton/monopole field, preserving
 the  $SO(5)$/$SU(N+1)$- symmetry of $S^4$/$\DC P^N$
and  providing  the system with a degenerate ground state \footnote{The $S0(5)$ symmetry of the particle on $S^4$
in the $SU(2)$ instanton  field has been observed for the first time in \cite{horv} and rediscovered by many authors.}.
Factorizing $S^7$ by $U(1)$, one can
get that  $\DC P^3$ is the $S^2$-fibration  over $S^4$,
which is well known from twistor theory.
It was mentioned in Ref. \cite{kn}  that the quantum Hall effect on
$\DC P^3$ should be
 connected with the quantum Hall effect on $S^4$ via this construction.

In this paper we establish an explicit correspondence
between classical mechanics and underlying Hall effects on $\DC P^3$ and $S^4$.
Namely, we reduce the Hamiltonian system
describing the motion of a particle on $\DC P^3$
in a constant magnetic field
(``the Landau problem on $\DC P^3$") to the Hamiltonian system, describing
the motion of particle on $S^4$ in
the $SU(2)$-instanton field (``four-dimensional Hall mechanics").
For this purpose, we  rewrite the Fubini-Study
 metric on $\DC P^3$ as follows:
\beq
g_{a\bar b}dz_ad\bar z_b=\frac{dzd\bar z}{1+z\bar z}-
\frac{(\bar z dz)(zd\bar z)}{(1+z\bar z)^2}=
\frac{dw_id\bar w_i}{(1+w\bar w)^2}+
\frac{(du+{\cal A})(d\bar u+\bar{\cal  A})}{(1+u\bar u)^2},
\label{map}\eeq
where
\beq
z_1=w_1u-\bar w_2,\quad z_2=w_2u+\bar w_1,\quad z_3=u,
\label{bt}\eeq
while
\beq
{\cal A}=\frac{(\bar w_1+w_2u)(udw_1-d\bar w_2) +(\bar w_2-w_1u)(udw_2+d\bar w_1)}{1+w\bar w}.
\label{A}\eeq
Here  $w_1,w_2$ and $u$ are  the conformal-flat
complex coordinates of  $S^4=\DH P^1$ and $S^2=\DC P^1$, while the
connection ${\cal A}$ defines the $SU(2)$ instanton field.
Hence, a particle on $\DC P^3$ could be
viewed as a particle on $S^4$ in the instanton field,
with non-fixed isospin (the $S^2$ sphere plays
the role of the internal space of the particle on $S^4$).
Then, fixing the isospin of the system,
 we reduce the particle on $\DC P^3$
to the four-dimensional Hall mechanics.
The phase space of the reduced system is  $T^*S^4\times S^2$. Surprisingly,
{\it performing a similar reduction for the Landau problem on $\DC P^3$
(i.e. a particle on $\DC P^3$ in a constant magnetic field)
we get a classically equivalent system. The
only difference with the previous system
is that the Hamiltonian is shifted by a constant.}

Notice   that the   effective theory of the Hall effect on $S^4$
is defined on fuzzy $\DC P^3$ \cite{Fabinger,ber},
 since in the large mass limit,
 when the upper Landau levels go to infinity, it
corresponds to the vanishing of the momenta of the system.
 In that case, the initial
phase space $T^*S^4\times S^2$ reduces to $\DC P^3\approx S^4\times S^2$.
 Clearly, the effective theory of the Hall effect on $\DC P^3$
 is formulated on the same space.

\vspace{5mm}

The complex projective space $\DC P^3$ could be equipped with the K\"ahler structure given by the Fubini-Study metric
  \begin{equation}
  g_{a \bar b}dz_ad{\bar z}_b =
\frac{\partial^2 \log (1+z\bar z)}{\partial z_a \partial{\bar z}_{b}}dz_ad{\bar z}_b,
  \end{equation}
and with the Poisson bracket
\begin{equation}
   \{ f,g\}_0 =
 i\frac{\partial  f}{\partial {\bar z}_a}g^{{\bar a}b}
\frac{\partial  g}{\partial z_b} -
i\frac{\partial g}{\partial z_b}g^{{\bar a}b}
\frac{\partial  f }{\partial {\bar z}_a},\quad
g^{{\bar a}b}
g_{b{\bar c}}=\delta^{\bar a}_{\bar c}.      \label{p0}
\end{equation}
The isometries of the K\"ahler structure form a $su(4)$ algebra
generated by the {\it holomorphic  Hamiltonian vector fields}
 \begin{equation}
{\vec V}_{\mu}=
    V_\mu^{a}(z)\frac
{\partial}{\partial z_a}+
{\bar V}_\mu^{\bar a}(\bar z) \frac{\partial}
{\partial \bar z_a}, \quad  [{\vec V}_{\mu},{\vec V}_{\nu}]=
C_{\mu \nu}^{\lambda}{\vec V}_{\lambda} ,
\label{vkillling}\end{equation}
where
\begin{equation}
{\vec V}_\mu=\{\ch_\mu, \}_0 , \quad
 \{ \ch_{\mu},\ch_{\nu}\}_0=
C_{\mu \nu}^{\lambda}
\ch_{\lambda},\quad
\frac{\partial^2 \ch_\mu}{\partial z_a \partial z_b} -
\Gamma^c_{ab}\frac{\partial \ch_\mu}{\partial z_c}=0.
\label{v}\end{equation}
The real functions $\ch_\mu$, called Killing potentials,
are given by the expressions
\beq
\ch_T= T^{\bar a b}\ch_{\bar a b}- {\rm tr}\;{\hat T}, \quad
\ch^1_a=\ch^-_a+\ch^+_a,\quad \ch^2_a=i(\ch^-_a-\ch^+_a),
\label{pkil}\eeq
where ${T}^{\bar a b}= {\bar{T^{\bar b a}}}$, and
\begin{equation}
\quad
\ch_{\bar a b}=\frac{z_a \bar z_b}{1+ z\bar z},\quad
\ch^-_a=\frac{z_a}{1+z\bar z},\quad
\ch^+_a=\frac{\bar z_a}{1+ z\bar z}.
\label{pkilling}\end{equation}
The algebra of $\ch_{\bar a b}, \ch^\pm_a$ reads
\beq
\begin{array}{c}
\{\ch_{{\bar a} b}, \ch_{\bar c d}\}_0=
i\delta_{\bar a d}\ch_{\bar b c}
-i\delta_{\bar c b}\ch_{\bar a d},\\
\{\ch^-_a, \ch^+_b\}_0=
i\delta_{\bar a b}(1-
{\rm tr}\;\ch_{\bar a b})+i\ch_{\bar a b},
\quad\{\ch^{\pm}_a, \ch^{\pm}_b\}_0=0,
\quad
\{\ch^{\pm}_a, \ch_{\bar b c}\}_0=
\mp i\ch^{\pm}_b\delta_{ a b}\quad .
\end{array}
\label{isoalg}\eeq

Let us define  on  $T^*\DC P^3$ the Hamiltonian system describing the motion of a free particle on $\DC P^3$
\begin{equation}
{\cal D}= g^{a \bar b}\pi_a{\bar \pi}_b \;,\quad
\{f,g\}=
\frac{\partial f}{\partial z_a}\frac{\partial g}{\partial \pi_a}-\frac{\partial g}{\partial \pi_a}
\frac{\partial f}{\partial\pi_a} + \quad c.\;c. \quad .
\label{ssB}\end{equation}
This system has a $su(4)$e symmetry defined by the
Noether's constants of motion
 \begin{equation}
 \begin{array}{c}
J_\mu=V_\mu^{a}\pi_a +
 {\bar V}_{\mu}^{\bar a} {\bar\pi}_{\bar a} :\quad
  \{{\cal D}, J_{\mu}\}=0, \\
J_{a\bar b}=-{i}z_b\pi_a+ i\bar\pi_b \bar z_a \;,\quad
iJ^{+}_a=\pi_a+\bar z_a(\bar z\bar\pi),\quad
-iJ^{-}_a=\bar\pi_a+ z_a(z\pi).
\end{array}
\label{jab}\end{equation}
In order to connect this system with Hall mechanics on $S^4$,
let us embed the seven-sphere $S^7$ in the eight-dimensional euclidean
 space $\DC^4=\DH^2$,
 parameterized by four complex (two quaternionic) coordinates
\beq
{\bf v}_i=v_i+jv_{i+1},\quad i=1,2,\quad { \bf v}_{1},\;{\bf v}_2\in \DH, \quad v_1,v_2,v_3,v_4\in \DC\quad .
\eeq
Then, let us consider a five-dimensional Euclidean
 space $\DR^5$ parameterized by the coordinates
\footnote{This map is known in quantum mechanics with the name of Hurwitz
transformation. It relates the eight-dimensional oscillator with a
five-dimensional Coulomb system with a $SU(2)$ Yang monopole (see, e.g.
\cite{hurwits}).}
\beq
{\bf w}=w_1+jw_2=2{\bf v}_1{\bf \bar v}_2,\quad x_5={\bf v}_1{\bf \bar v}_1-{\bf v}_2{\bf \bar v}_2,
\qquad {\bf w}\in \DH,\;x_5\in \DR.
\label{hurwitz}\eeq
It is seen that (\ref{hurwitz})
is invariant under the right action of a $SU(2)$ group
\beq
{\bf v}_i\to {\bf v}_i {\bf  g},\quad {\rm where}\;\; {\bf g}\in\DH, \quad {\bf g}{\bf \bar g}=1
\eeq
Now, defining in $\DH^2$ the seven-sphere $S^7$, by the $SU(2)$-invariant constraint
${\bf v}_i{\bf\bar v}_i=1$,
we can get the four-sphere $S^4$
embedded in $\DR^5$:
${\bf w\bar w} +x^2_5=1$.
 Taking into account that ${\bf g}$ defines a
three-sphere $S^3$, we get the second Hopf map, $S^7/S^3=S^4$.
The complex projective space $\DC P^3$ is defined as $S^7/U(1)$,
while the inhomogeneous coordinates $z_a$ appearing in the
Fubini-Study metric of $\DC P^3$,
 are related with the coordinates of $\DC^4$ as follows:
 $z_a=\lambda_a/\lambda_4$, $a=1,2,3$.
The expressions (\ref{hurwitz}) defining $S^4$
 are invariant under $U(1)$-factorization,  while  $S^3/U(1)=S^2$.
Thus, we  arrive to the conclusion that  $\DC P^3$ is the $S^2$-fibration over  $S^4=\DH P^1$.
The expressions for $z_a$
 yield the following definition of the coordinates of $S^4$:
\beq
w_1=\frac{\bar z_2 +z_1\bar z_3}{1+z_3\bar z_3},\quad
w_2=\frac{ z_2\bar z_3 -\bar z_1}{1+z_3\bar z_3}.
\label{wz}\eeq
Choosing $z_3$ as a local coordinate of  $S^2=\DC P^1$,
\beq
u=z_3\;,
\label{uz3}\eeq
we  get the expressions (\ref{bt}).

Now, let us  define the canonical transformation of (\ref{ssB}),
 completing  the coordinate transformation (\ref{bt})
by the following transformation of momenta
\beq
\pi_1=\frac{\bar u p_1-\bar p_2}{1+u\bar u},\quad
\pi_2=\frac{\bar u p_2+ p_1}{1+u\bar u},\quad
\pi_3= p_u+\frac{\bar  p_2 w_1-\bar p_1 w_2 -
\bar u (w_1p_1+w_2p_2)}{1+u\bar u}.
\label{pcan}\eeq
The transformed  Hamiltonian system
reads as follows
\beq
  {\cal D}_0=(1+w\bar w)^2P_i\bar P_i +(1+u\bar u)^2p_u\bar p_u\; ,\quad
  \{f,g\}=
\frac{\partial f}{\partial w_i}\frac{\partial g}{\partial p_i}-\frac{\partial f}{\partial p_i}
\frac{\partial g}{\partial w_i} +
\frac{\partial f}{\partial u}\frac{\partial g}{\partial p_u}-\frac{\partial f}{\partial p_u}
\frac{\partial g}{\partial u}\quad +\quad c.\;c. \; .
\label{htrans}\eeq
Here we introduced the covariant momenta
\beq
P_1=p_1-i\frac{\bar w_1 }{1+w\bar w}I_1-\frac{w_2}{1+w\bar w}I_+,
\quad P_2=p_2-i\frac{\bar w_2 }{1+w\bar w}I_1+\frac{w_1}{1+w\bar w}I_+,
\label{Pcov}\eeq
and  the $su(2)$ generators  $I_\pm, I_1$ defining the isometries of $S^2$:
\beq\begin{array}{c}
I_1=-i(p_u u-\bar p_u \bar u), \quad I_-=p_u+\bar u^2\bar p_{\bar u},
\quad  I_+=\bar p_{\bar u}+ u^2 p_u\\
\{I_\pm, I_1\}= \mp iI_\pm,\quad \{I_+, I_-\}= 2iI_1.
\end{array}
\label{Kiso}\eeq
The nonvanishing Poisson brackets between $P_i$, $w_i$ are given by the
following relations (and their complex conjugates)
\beq
\{w_i,P_j\}=\delta_{ij},\quad \{ P_1, P_2\}=-\frac{2I_+}{(1+w\bar w)^2},
\quad\{ P_i,\bar  P_j\}=-i\frac{2I_1\delta_{ij}}{(1+w\bar w)^2}.
\label{instant}\eeq
The expressions in the r.h.s. define the strength of a
homogeneous
 $SU(2)$ instanton,  written in terms of conformal-flat
coordinates of $S^4=\DH P^1$. Hence, the first part of the Hamiltonian, i.e.
${\cal D}_4=(1+w\bar w)^2P_i\bar P_i$, describes a
 particle on the four-dimensional sphere in the field of a $SU(2)$ instanton.\\
The generators of its  symmetry
 algebra, i.e. $so(5)$, are  connected with the symmetry generators of $\DC P^4$
as follows:
\beq
\begin{array}{c}
{L}_{1\bar 1}=  {J}_{2,\bar 2}+{J}_{3,\bar 3} =i(\overline{{w}_{1}} \, \overline{{p}_{1}}-{w}_{1} \,
{p}_{1})-{I}_{1},\\
{L}_{2\bar 2}= {J}_{1,\bar 1}+{J}_{3,\bar 3} =i(\overline{{w}_{2}} \, \overline{{p}_{2}}-{w}_{2} \,
{p}_{2})-{I}_{1}\\
{L}^+_{1\bar 2}=-i \, {J}_{1,\bar 2}={w}_{2} \, {p}_{1}-\overline{{w}_{1}}
\,
\overline{{p}_{2}},\\
{L}^+_{12}=i \, {J}_{3}^{-}=\overline{{w}_{2}} \,
{p}_{1}-\overline{{w}_{1}} \, {p}_{2}-{I}_{+}\\
{L}^+_{1}=-i \, \left ( {J}_{1,\bar 3}+{J}_{2}^{-}\right ) =\left (
1-{w}\,\overline{{w}}\right )  \, {p}_{1}+\overline{{w}_{1}} \,
\left ( \overline{{w}} \,
\overline{{p}}+{w} \, {p}\right )
+{w}_{2} \, {I}_{+}+i\overline{{w}_{1}} \,
{I}_{1}\\
{L}^+_{2}=-i \, \left ( {J}_{2,\bar 3}-{J}_{1}^{-}\right ) =\left (
1-{w} \, \overline{{w}}\right )  \, {p}_{2}+\overline{{w}_{2}} \,
\left ( \overline{{w}} \,
\overline{{p}}+{w} \, {p}\right )
 -{w}_{1} \, {I}_{+}+i\overline{{w}_{2}} \,
{I}_{1}
\end{array}
\label{isometries}
\eeq
Here $L^{\pm}_{i\bar j}, L^{\pm}_{12}$ denote the generators of $so(4)$
rotations,
and $L^\pm_a $ are the  generators of translations.

The  Poisson brackets between $P_a$ and $u,\bar u, p_u,\bar p_u$
 are  defined by the following nonzero relations and their complex conjugates:
\beq \{ P_i, p_u\}=-\frac{ \overline{{w}_{i}}+2 \epsilon_{ij}\, {w}_{j} \, u}
{1+\overline{{w}}{w}}p_u,\;\; \{ P_i,\bar p_u\}=\frac{\overline{w}_i{\bar p_
u}}{1+\overline{w}{w}},  \;\; \eeq \beq \left\{ P_{i},u\right\}  =\frac{ \left
( \overline{w_{i}}+\epsilon_{ij}w_{j} \, u\right )u }{1+\overline{w} w}\;\;,
\left\{ \overline{P_{i}},u\right\}  =\frac{ \epsilon_{ij}\overline{w_{j}}-w_{i}
\, u}{1+\overline{w}w}. \label{Ppbrac}\eeq The second part of the Hamiltonian
defines the motion of a free particle on the two-sphere. It could be
represented as a Casimir of $SU(2)$ \beq {\cal D}_{S^2}=(1+u\bar u)^2p_u\bar
p_u=I_+I_-+I_1^2\equiv I^2. \label{D2}\eeq It commutes with the Hamiltonian
${\cal D}_0$,
 as well as with  $I_1, I_\pm$ and  $P_i, w_i$:
\beq
\{{\cal D}_{\DC P^3}, I^2\}=
\{P_i,{ I}^2\}_B= \{w_i,{ I}^2\}_B=
\{{\cal I}_1,{ I}^2\}_B=\{{\cal I}_\pm ,{I}^2\}_B = 0.
\label{nen}\eeq
Hence, we can perform a Hamiltonian
reduction by the action of the generator  ${\cal D}_2$, which
reduces the initial twelve-dimensional phase space $T_*\DC P^3$ to a
ten-dimensional one. The relations (\ref{nen})
allow us to  parameterize the reduced ten-dimensional phase space
in term of the coordinates
 $P_i, w_i, I_\pm, I_1$,
where the latter obey the relation
\beq
I_+I_- +I^2_1\equiv I^2=const\;.
\eeq
Thus, the reduced phase space is $T^*S^4 \times S^2$, where
$S^2$ is  the internal space of the instanton.

Let us collect the
 whole set of non-zero expressions  defining  the Poisson brackets
on  $T_*S^4 \times S^2$: \beq \{w_i, P_j\}=\delta_{ij},\;\;\{ P_1,
P_2\}=-\frac{2I_+}{(1+w\bar w)^2}, \;\; \{ P_i,\bar
P_j\}=-i\frac{2I_1\delta_{ij}}{(1+w\bar w)^2}, \label{Pb1} \eeq \beq \left\{
P_{i},I_{1}\right\} =i\frac{\epsilon_{ij}w_{j} \, I_{+}}{1+ w\overline{w}}
\quad \left\{ P_{i},I_{+}\right \} =\frac{\overline{w_{i}} \, I_{+}}{1+
\overline{w} w},\quad \left \{{P _{i}},I_{-}\right\} = -\frac{ \overline{w_{i}}
\, I_{-}+ 2i\epsilon_{ij}{w_{j}}\, I_{1} }{1+\overline{w} w} \label{Pb2}\eeq
\beq \{I_+,I_-\}=2iI_1,\quad \{I_\pm, I_1\}=\mp iI_\pm \; . \label{Pb3} \eeq
The reduced Hamiltonian is $ {\cal D}_{\DC P^3}^{red}=(1+w\bar w)^2 P\bar P
+I^2$. So, the Hamiltonian of four-dimensional Hall mechanics  is connected
with the Hamiltonian of a particle on $\DC P^3$ as follows:
\beq {\cal D}_{S^4}={\cal D}_{\DC P^3}^{\rm red}-I^2 \quad(>0). \eeq
This yields an
intuitive explanation of the degeneracy in the ground
 state of the quantum  Hall mechanics on $S^4$. Indeed,
 since the l.h.s. is positive,
the ground state of the quantum Hall
 mechanics on $S^4$ corresponds to the excited state of a
 particle on $\DC P^3$, which is a degenerate one.
 On the other hand, the ground state of a particle on $\DC P^3$ can be
 reduced to the $S^4$ Hall mechanics when $I=0$, which corresponds to a free particle on $S^4$. In the case
  $I\to\infty$ corresponding to the thermodynamic  limit of quantum Hall effect
on $S^4$,  we get that the ground state on $S^4$
 corresponds to the infinitely higher state of a particle on $\DC P^3$.

The inclusion of a constant magnetic field,
 i.e. the consideration of the Landau problem on $\DC P^3$,
 provides the particle with a degenerate ground state.
 This is the reason why Karabali and Nair
 were able to observe the Hall effect on $\DC P^N$ \cite{kn}.
Let us consider the classical
 correspondence between the Landau
problem on $\DC P^3$ and the Hall mechanics on $S^4$,
in order to clarify which modifications
in the above picture are induced by the inclusion of a magnetic field.

For considering the Landau problem on $\DC P^3$,
 we   modify the initial Hamiltonian system (\ref{ssB}) as follows:
\begin{equation}
 {\cal D}_{\DC P^3}= g^{a \bar b}\pi_a{\bar \pi}_b,\quad \{f, g \}_B=\{f, g\}+iBg_{a \bar b}\left(
 \frac{\partial f}{\partial \pi_a}
\frac{\partial g}{\partial \bar \pi_b}
-\frac{\partial g}{\partial  \pi_a}
\frac{\partial f}{\partial \bar \pi_b}\right).
\label{ssB1}\end{equation}
The isometries of
$\DC P^3$ define the
Noether's constants of motion
 \begin{equation}
{\cal J}_{\mu}\equiv J_\mu+ B\ch_\mu=V_\mu^{a}\pi_a +
 {\bar V}_{\mu}^{\bar a} {\bar\pi}_{\bar a} +B\ch_\mu :
  \{{\cal D}_{\DC P^3}, {\cal J}_{\mu}\}_B=0, \;
  \{{\cal J}_\mu,{\cal J}_\nu\}_B=C_{\mu\nu}^\lambda {\cal J}_\lambda .
\end{equation}
The vector fields
 generated by ${\cal J}_\mu$ are independent on $B$
\begin{equation}
{\vec V}_{p}=V^a(z)\frac{\partial}{\partial z_a}-V^a_{,b}\pi_a
\frac{\partial}{\partial \pi_a}+
{\bar V}^a(\bar z)\frac{\partial}{\partial\bar z^a}
-{\bar V}^a_{,\bar b}\bar\pi_a
\frac{\partial}{\partial \bar\pi_a}\quad ,
\label{vkillingB}\end{equation}
i.e.  the inclusion of a
 constant magnetic field preserves
the whole symmetry algebra of
a particle  on $\DC P^3$.

Passing to the coordinates (\ref{bt}) and momenta (\ref{Pcov})
we get the Poisson brackets defined by the nonzero relations given by
(\ref{Ppbrac}) and
\beq
\{p_u\overline{p}_u\}_B=\frac{iB}{(1+u\overline{u})^2},
\label{pupub}\eeq
\beq
\{w_i, P_j\}_B=\delta_{ij},\quad \{ P_1, P_2\}_B=-\frac{2{\cal I}_+}{(1+w\bar w)^2}, \quad
\{ P_i,\bar  P_j\}_B=-i\frac{2{\cal I}_1\delta_{ij}}{(1+w\bar w)^2}.
\label{instantB}\eeq
where
${\cal I}_\pm, {\cal I}_1$ are defined by the expressions
\beq
{\cal I}_1=I_1+\frac B2\frac{1-u\bar u}{1+u\bar u},
 \quad {\cal I}_-=I_- -B\frac{i\bar u}{1+u\bar u},
\quad  {\cal I}_+= I_+ +B\frac{i u}{1+u\bar u}\;
\label{Kiso1}
\eeq
Notice that the expressions (\ref{instantB}) are similar to
(\ref{instant}) and
the generators (\ref{Kiso1})  form,
 with respect to the new Poisson brackets, the $su(2)$ algebra
\beq
\{{\cal I}_\pm, {\cal I}_1\}_B= \mp i{\cal I}_\pm,\quad
 \{{\cal I}_+, {\cal I}_-\}= 2i{\cal I}_1.
\eeq
It is clear that these generators define the isometries
of the ``internal" two-dimensional sphere  with a magnetic monopole
located in the center.

Once again, as in the absence of magnetic field, we can reduce the
initial system by the Casimir of the $SU(2)$ group
 \beq
{\cal I}^2\equiv {\cal I}_1^2 + {\cal I}_+ {\cal I}_-
={\cal D}_{S^2} +B^2/4,\quad \Rightarrow \quad
{\cal I}\geq B/2\; .
 \label{casB}\eeq
In order to perform the Hamiltonian reduction,
we have to fix the value of ${\cal I}^2$,
and then factorize by the action of vector field $\{{\cal I}^2,\;\}_B $.

The coordinates (\ref{wz}), (\ref{Pcov}) commute with the Casimir (\ref{casB}),
\beq
\{P_i,{\cal I}^2\}_B= \{w_i,{\cal I}^2\}_B=
\{{\cal I}_1,{\cal I}^2\}_B=\{{\cal I}_\pm ,{\cal I}^2\}_B = 0.
\eeq
Hence, as we did above, we can choose
$P_i$, $w_i$, and ${\cal I}_\pm$ as the coordinates of the reduced,
 ten-dimensional phase space.

The coordinates ${\cal I}_\pm$, ${\cal I}_1$
obey the condition
\beq
{\cal I}^2_1+{\cal I}_-{\cal I}_+={\cal I}^2=const.
\eeq
The resulting Poisson brackets are defined by the expressions (\ref{Pb1}),
(\ref{Pb2}),(\ref{Pb3}), where $I_1, I_\pm$
are replaced by ${\cal I}_\pm,{\cal I}_1$.

Hence, the Landau problem on $\DC P^3$  reduces to
the Hall mechanics on $S^4$
whose Hamiltonian
is defined by the expression
\beq
{\cal D}_{S^4}={\cal D}_{\DC P^3}^{red}-{\cal I}^2+B^2/4,
\eeq
where
\beq
{\cal I}\geq B/2.
\eeq
Notice that upon quantization we must replace ${\cal I}^2$ by
 ${\cal I}({\cal I}+1)$ and require that both ${\cal I}$ and
$B$  take (half)integer values (since we assume unit radii for the spheres,
this means, that the ``monopole number" obeys a Dirac quantization rule).

We get the following surprising result:
the Landau problem on $\DC P^3$ yields,
actually, the same  four-dimensional system on
$\DC P^3$ as  a free particle.
The only difference between the resulting systems
is that the isospin of the reduced Landau problem
has a lower bound ${\cal I}> B/2$,
 while the reduced free particle could be equipped with any isospin;
the ground state  of the Landau problem on $\DC P^3$ corresponds to the
excited state of four-dimensional Hall mechanics.\\

In the  higher-dimensional Hall effect an
essential role is played by the inclusion of a potential field, breaking
the initial symmetry of $S^4$ and $\DC P^N$(i.e.
 $SO(5)$  and $SU(N+1)$, respectively)
to $SO(4)$ and $SU(N)$
\beq
{\cal H}_{S^4}={\cal D}_{S^4} +V(w\bar w)\;, \quad
{\cal H}_{\DC P^N}={\cal D}_{\DC P^N} +V(z\bar z)\;.
\eeq
The inclusion of a potential
collects the particle on the  edges
of $S^4$ and $\DC P^N$, which are, respectively,
three- and $(2N-1)-$ dimensional.
The relations between the coordinates of $\DC P^3$ and $S^4$ read
\beq
1+z\bar z=(1+w\bar w)(1+ u\bar u), \quad u=z_3,
\eeq
so that the system on $\DC P^3$ with a $SU(3)$-invariant  potential
 cannot be reduced to the Hall mechanics of a system on $S^4$ in the instanton field.
For example, the oscillator  on $\DC P^3$  specified by the potential $V=\omega^2 z\bar z$
 which has the hidden symmetry \cite{bn}
cannot be reduced to the system on $S^4$ in the $SU(2)$ instanton field.
On the other hand, one can believe
  that the hidden symmetry (or, at least, the exact solvability) of the oscillator on $S^4$ with the potential
   $V=\omega^2 w\bar w/(1-w\bar w)^2$ \cite{higgs}, would  be preserved after inclusion of the instanton field.
If so, one can further reduce the system to the three-dimensional
Coulomb-like system (see \cite{np}), in order to obtain the exactly-solvable generalization of the
Coulomb system with a non-Abelian monopole.

In the large mass limit the upper energy levels of the Hall mechanics on $S^4$ run  to infinity,
while the lowest Landau levels are
described by the non-commutative mechanics specified by the  Hamiltonian ${\cal H}_{NC}=V$,
with the phase space $\DC P^3$ \cite{4h,ber,kn}. So, the Hall effect on $S^4$ corresponds to
the Hall effect on $\DC P^3$ when the potential field in the latter one breaks
the rotational $SU(3)$ symmetry of the system; the infinite number of internal degrees of freedom arising
in the Hall effect on $S^4$ corresponds to the spatial degrees of freedom in $\DC P^3$.
Notice, that  the  noncommutative mechanics on $\DC P^1\approx S^2$ in a constant magnetic field reduces, at some
``critical point",  to the one with the
phase space $\DC P^1$ \cite{nc}. There is no doubt, that a similar mechanism would exist for the noncommutative
mechanics on $\DC P^3$, i.e. the  large mass limit of the  Hall  mechanics on $S^4$
could be viewed as the ``critical point" of noncommutative mechanics on $\DC P^3$ (with a finite mass)
in a constant  magnetic field.

Finally, a few remarks about supersymmetrization are in order. A particle on $\DC P^3$
could be easily endowed with ${\cal N}=4$ supersymmetry, due to the
K\"ahler structure of the configuration space \cite{n4}. Moreover,
supersymmetry can be preserved  upon inclusion of a constant magnetic
field, since $\DC P^3$ has a constant curvature. Considering $\DC P^3$ as a
$S^2$-fibration over $S^4$, we can interpret the constructed
supersymmetric system on $\DC P^3$ as a supersymmetric extension of
four-dimensional Hall mechanics, with some   additional
degree of freedom. The extra degree of
freedom is introduced when extending the above presented hamiltonian reduction to the
appropriate super-Hamiltonian one. Performing this step, one could get a
well-defined  ${\cal N}=4$ supersymmetric four-dimensional Hall mechanics.
On the other hand, in the large mass limit, both the Landau problem on $\DC P^3$
and the Hall mechanics on $S^2$ result in systems on $\DC P^3$, which
could equipped with ${\cal N}=4$ supersymmetry by an appropriate choice for the
Hamiltonian ${\cal H}_{NC}=g^{\bar a b}\partial_{\bar a}{\bar U}(\bar
z)\partial_b U(z)$.
\\

{\large Acknowledgements}
The work of S.B. was supported in part
by the European Community's Human Potential
Programme under
contract HPRN-CT-2000-00131 Quantum Spacetime,
the INTAS-00-0254 grant and the
NATO Collaborative Linkage Grant PST.CLG.979389.
P.-Y.C. is supported by a postdoctoral fellowship of the Ecole Polytechnique.
The work of A.N. was supported by grants INTAS 00-00262  and ANSEF  PS81.
A.N. thanks INFN-LNF
for hospitality during the completion of  this work. A stay of A.N. at the Joint Institute for Nuclear Research (Dubna)
 is also acknowledged.

\end{document}